# Adversarial attacks on an optical neural network

Shuming Jiao[1], Ziwei Song[2], and Shuiying Xiang[2]
1. Peng Cheng Laboratory, Shenzhen, Guangdong, China
2. State Key Laboratory of Integrated Service Networks, Xidian University, Xi'an, China

**Abstract**
*Adversarial attacks have been extensively investigated for machine learning systems including deep learning in the digital domain. However, the adversarial attacks on optical neural networks (ONN) have been seldom considered previously. In this work, we first construct an accurate image classifier with an ONN using a mesh of interconnected Mach–Zehnder interferometers (MZI). Then a corresponding adversarial attack scheme is proposed for the first time. The attacked images are visually very similar to the original ones but the ONN system becomes malfunctioned and generates wrong classification results in most time. The results indicate that adversarial attack is also a significant issue for optical machine learning systems.*

In the modern information era with massive data, efficient computer technology is always an urgent demand. In recent years, optical computing and optical neural network (ONN) have received much research attention [1,2]. Compared with electronic computing, optical computing has potential advantages of high-speed parallel processing and low power consumption.

Various ONN schemes in analogy to digital deep learning neural network have been proposed such as cascaded Mach–Zehnder interferometer (MZI) network [3-6], optical diffractive neural network [7-9] and wavelength multiplexing network [10]. Among them, an ONN with cascade MZIs has advantages of compact system size and good compatibility with silicon photonics. An ONN with cascaded MZIs can be used for machine learning tasks such as voice and image classification. For digital machine learning algorithms including deep learning, adversarial attacks have been extensively investigated [11-13]. They pose general and serious threats to machine learning and the revealed system vulnerability has profound implications. As an interesting phenomenon, adversarial attacks refer to perturbing input images subtly with minor changes for intentionally generating wrong classification results by "fooling" the machine learning system. However, the adversarial attack on an ONN was seldom noticed and considered in the past. In some previous works [14-16], optical imaging systems are used to generate or prevent adversarial attack for a digital deep learning system while the attacking target is not an optical machine learning system. In this paper, adversarial attacks are performed on a MZI mesh system for the first time, as a typical type of ONN.

In an ONN with cascaded MZIs, each MZI unit has two input ports and two output ports and a MZI mesh consists of many interconnected MZIs. The entire MZI mesh can model a vector-matrix multiplication operation. For example, an input vector consisting of 4 elements is multiplied with a 4×4 matrix and the output vector consists of 4 elements. Then the cascaded

MZI system will have 4 input ports and 4 output ports respectively. The intensities of coherent light signals at different corresponding input or output ports correspond to the vector element values. Each individual MZI unit has two phase shifters and their phase values can be flexibly adjusted. A mesh with a sufficient number of MZI units can model an arbitrary weighting matrix if all the phase values are assigned with optimized values.

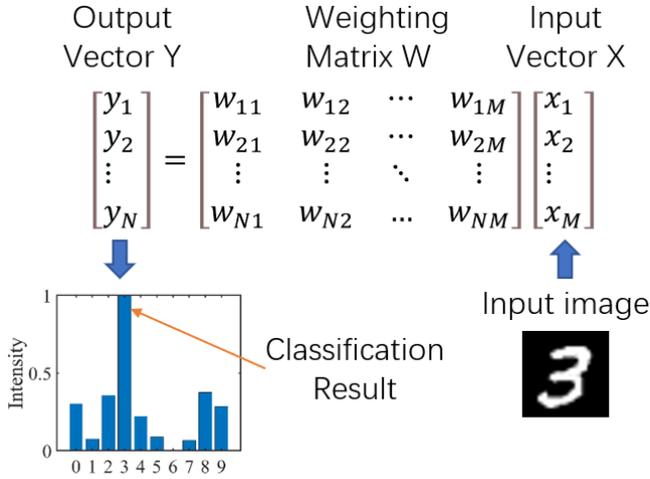

Fig. 1 An image classifier by vector-matrix multiplication

A linear classifier for sorting images consisting of M pixels into N categories can be implemented based on vector-matrix multiplication, shown in Fig. 1. If the input vector X represents pixel values of the input image, the classification result can be indicated by the maximum value in the output vector Y. Based on many training samples (different images with known classification results), the element values of weighting matrix W can be optimized. Each matrix element can be updated iteratively by the following formulas (1) and (2) for each $i\ (1 \leq i \leq N)$ and $j\ (1 \leq j \leq M)$ with a learning rate $\mu$. The optimization is based on gradient decent and softmax function.

$$w_{ij} \leftarrow w_{ij} - \mu(T_i - y_i)x_j \quad (1)$$

$$T_i = \begin{cases} \frac{\exp(y_i)}{\sum_{i=1}^{N} \exp(y_i)}, & X: not\ ith\ category \\ \frac{\exp(y_i)}{\sum_{i=1}^{N} \exp(y_i)} - 1, & X: ith\ category \end{cases} \quad (2)$$

After training, all the elements in W can be further normalized to [0 1] to ensure that all the output vector element values are positive if the input values are positive. For a given image with unknown category, the vector-matrix multiplication operation with optimized W can yield the classification result.

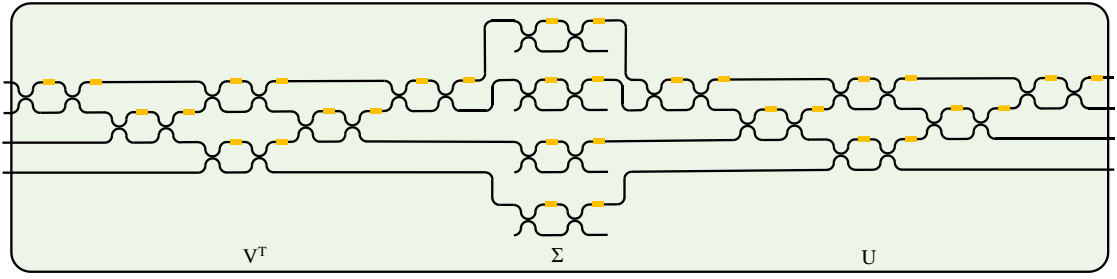

Fig. 2 Basic block of MZI network for a $4 \times 4$ submatrix

The image classifier can be implemented by an ONN with cascaded MZIs in the following way. First, a basic block of MZI network can be constructed for a $4 \times 4$ submatrix. An arbitrary matrix can be decomposed into three matrices by singular value decomposition (SVD): unitary matrix U, diagonal matrix $\Sigma$, and the other unitary matrix $V$. As shown in Fig. 2, a MZI network is composed of three parts to implement three matrices respectively. There is a total of 16 MZI units in the MZI mesh in which each MZI can be reconfigured individually. The phase values of each MZI in a $4 \times 4$ MZI network can be optimized based on three successive multiplications [17,18].

$$Y_{N \times 1} = W_{N \times M} \cdot X_{M \times 1} = \begin{bmatrix} w_{11} & w_{12} & \cdots & w_{1M} \\ w_{21} & w_{22} & \cdots & w_{2M} \\ \cdots \\ w_{N1} & w_{N2} & \cdots & w_{NM} \end{bmatrix} \begin{bmatrix} x_1 \\ x_2 \\ \vdots \\ x_M \end{bmatrix} = \begin{bmatrix} w_{11}x_1 + w_{12}x_2 + \cdots + w_{1M}x_M \\ w_{21}x_1 + w_{22}x_2 + \cdots + w_{2M}x_M \\ \vdots \\ w_{N1}x_1 + w_{N2}x_2 + \cdots + w_{NM}x_N \end{bmatrix} = \begin{bmatrix} y_1 \\ y_2 \\ \vdots \\ y_N \end{bmatrix}$$

$$\underbrace{(w_{N1}x_1 + \cdots + w_{N4}x_4)}_{y_{N1}^{(1)}} + \underbrace{(w_{N5}x_5 + \cdots + w_{N8}x_8)}_{y_{N1}^{(2)}} + \underbrace{(w_{N9}x_9 + \cdots + w_{N12}x_{12})}_{y_{N1}^{(3)}} + \underbrace{(w_{N13}x_{13} + \cdots + w_{N16}x_{16})}_{y_{N1}^{(4)}} + \cdots + w_{NM}x_M$$

$$\begin{bmatrix} w_{N1} & w_{N2} & w_{N3} & w_{N4} \\ w_{N5} & w_{N6} & w_{N7} & w_{N8} \\ w_{N9} & w_{N10} & w_{N11} & w_{N12} \\ w_{N13} & w_{N14} & w_{N15} & w_{N16} \end{bmatrix} \begin{bmatrix} x_1 \\ x_2 \\ x_3 \\ x_4 \end{bmatrix} \quad \begin{bmatrix} w_{N1} & w_{N2} & w_{N3} & w_{N4} \\ w_{N5} & w_{N6} & w_{N7} & w_{N8} \\ w_{N9} & w_{N10} & w_{N11} & w_{N12} \\ w_{N13} & w_{N14} & w_{N15} & w_{N16} \end{bmatrix} \begin{bmatrix} x_5 \\ x_6 \\ x_7 \\ x_8 \end{bmatrix} \quad \begin{bmatrix} w_{N1} & w_{N2} & w_{N3} & w_{N4} \\ w_{N5} & w_{N6} & w_{N7} & w_{N8} \\ w_{N9} & w_{N10} & w_{N11} & w_{N12} \\ w_{N13} & w_{N14} & w_{N15} & w_{N16} \end{bmatrix} \begin{bmatrix} x_9 \\ x_{10} \\ x_{11} \\ x_{12} \end{bmatrix} \quad \begin{bmatrix} w_{N1} & w_{N2} & w_{N3} & w_{N4} \\ w_{N5} & w_{N6} & w_{N7} & w_{N8} \\ w_{N9} & w_{N10} & w_{N11} & w_{N12} \\ w_{N13} & w_{N14} & w_{N15} & w_{N16} \end{bmatrix} \begin{bmatrix} x_{13} \\ x_{14} \\ x_{15} \\ x_{16} \end{bmatrix}$$

$$W_N^1 \qquad W_N^1 \qquad W_N^1 \qquad W_N^1$$

Fig. 3 Matrix decomposition into $4 \times 4$ submatrices for vector-matrix multiplication with a basic block of MZI networks by temporal multiplexing

The vector-matrix multiplication for the image classifier above can be implemented by a temporal multiplexing of many basic blocks of $4 \times 4$ MZI networks, and the decomposition of calculation tasks is shown in Fig. 3. Each row of the matrix $W_{N \times M}$ is divided into groups of every 16 elements and each group is rearranged as a $4 \times 4$ matrix. Thus, the matrix $W_{N \times M}$ is decomposed into a total of $(M/16)*N$ submatrices and each row corresponds to M/16=49 submatrices. That means the $4 \times 4$ MZI network will be used $(M/16)*N$ times by reconfiguring different phase values. As shown in Fig. 3, each submatrix receives a vector of four input elements at each time. The vector-matrix multiplication results of the $4 \times 4$ submatrices with four different sets of $4 \times 1$ input vectors are accumulated to obtain one of the elements in the output vector $y_N$ corresponding to the entire Nth row in the matrix, i.e., $y_N = y_{N1}^{(1)} + y_{N1}^{(2)} + y_{N1}^{(3)} + y_{N1}^{(4)} + \cdots + y_{N(M/16)}^{(1)} + y_{N(M/16)}^{(2)} + y_{N(M/16)}^{(3)} + y_{N(M/16)}^{(4)}$. For other elements in the output vector, the calculation processes are the same as described above. Finally, the output vector $Y = [y_1, y_2, \ldots y_N]^T$ can be obtained by a cascaded MZI system.

Our proposed corresponding adversarial attack scheme is described below. It shall be noticed that the ONN system remains fixed and unchanged in the whole process. Only the input test image is modified at a minimal level. As stated above, the maximum element in the output vector Y indicates which one of the N categories the input vector (or the input image) belongs to. It is assumed that X is originally correctly identified as the kth category by this system ($y_k$ is maximum). The objective is to find an attacked image visually similar to the original image but it will be mis-identified as another category different from kth one. In the original output vector, it is assumed that the second largest element is $y_l$. $\bar{X} = [\bar{x}_1 \, \bar{x}_2 \, \ldots \, \bar{x}_M]^T$ denotes the attacked image and the objective function $f(\bar{x}_m)$ $(1 \leq m \leq M)$ for minimization is given by Equation (3) where $\alpha'$, $\beta'$ and $\gamma$ are weighting coefficients. The first term is to suppress $y_k$ corresponding to the original correct category, the second term is to boost $y_l$ corresponding to a wrong category and the third term is to ensure that the attacked image is still similar to the original one.

$$f(\bar{x}_m) = \sum_{i=1}^{M}(\alpha' w_{lm} \bar{x}_m - \beta' w_{km} \bar{x}_m + \gamma |\bar{x}_m - x_m|^2) \quad (3)$$

An optimized solution of $\bar{X}$ can be obtained by gradient descent and minimization of mean square error with formulas (4) and (5) where $\mu'$ is the learning rate. The elements in $\bar{X}$ are initially set to be random before iterative optimization.

$$\frac{df}{d\bar{x}_m} = \alpha' w_{lm} - \beta' w_{km} + 2\gamma(\bar{x}_m - x_m) \quad (4)$$

$$x_m \leftarrow x_m - \mu' \frac{df}{d\bar{x}_m} \quad (5)$$

In this scheme, the classification result after attacking is conventionally not freely controlled. Alternatively, $y_l$ can be selected based on the intended category instead of the second largest element. In this case, the scheme is referred to as "selective attack".

The ONN system is first employed to classify number digit images (MNIST dataset [19]) and fashion product images (Fashion-MNIST dataset [20]). The number of pixels in each input image is M=28×28=784 and the number of categories is N=10. Therefore the size for the weighting matrix will be 10×784. The parameters are set as $\alpha = 1$, $\beta = 1$ and $\mu = 0.0001$. The number of iterations is 300. Totally 10000 samples are used to train the system and then another 10000 different images are used for testing. A classification accuracy of 91.3% (for MNIST) and 83.5% (for Fashion-MNIST) can be achieved.

Then the adversarial attacks described above are performed on 200 random test images of number digits (or fashion products) that are already correctly classified. The parameters are set as $\alpha' = 1$, $\beta' = 1$, $\gamma = 2$ and $\mu = 0.01$. The number of iterations is 300. Finally, 84% (for number digit) and 88% (for fashion product) of the attacked images become mis-classified. Some examples of successful attacked results are shown in Fig. 4

| Original image | | Attacked image | | Original image | | Attacked image | |
|---|---|---|---|---|---|---|---|
| 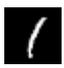 | 1 | 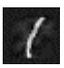 | 8 | 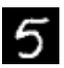 | 5 | 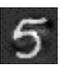 | 9 |
| 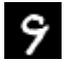 | 9 | 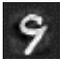 | 3 | 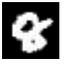 | 8 | 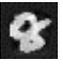 | 9 |
| 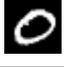 | 0 | 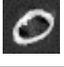 | 5 | 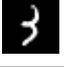 | 3 | 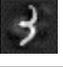 | 4 |
| 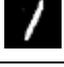 | 1 | 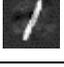 | 2 | 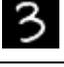 | 3 | 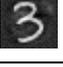 | 5 |
| 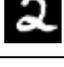 | 2 | 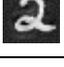 | 3 | 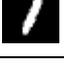 | 1 | 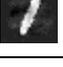 | 2 |
| 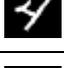 | 4 | 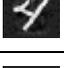 | 9 | 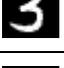 | 3 | 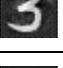 | 5 |
| 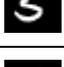 | 5 | 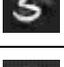 | 8 | 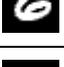 | 6 | 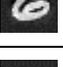 | 2 |
| 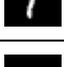 | 1 | 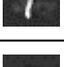 | 2 | 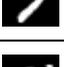 | 7 | 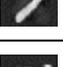 | 9 |
| 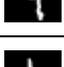 | 9 | 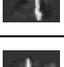 | 7 | 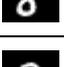 | 8 | 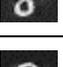 | 5 |
| 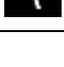 | 1 | 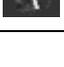 | 5 | 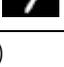 | 9 | 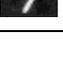 | 7 |

(a)

| Original image | | Attacked image | | Original image | | Attacked image | |
|---|---|---|---|---|---|---|---|
| 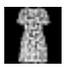 | dress | 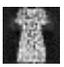 | T-shirt | 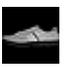 | sneaker | 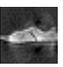 | sandal |
| 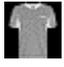 | T-shirt | 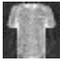 | shirt | 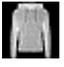 | pullover | 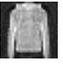 | coat |
| 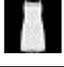 | dress | 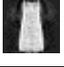 | T-shirt | 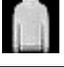 | pullover | 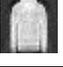 | coat |
| 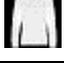 | pullover | 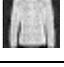 | shirt | 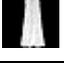 | dress | 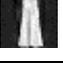 | trousers |
| 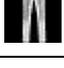 | trousers | 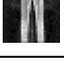 | T-shirt | 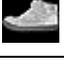 | sneaker | 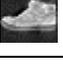 | boot |
| 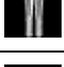 | trousers | 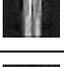 | dress | 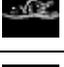 | sandal | 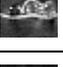 | sneaker |
| 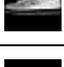 | sneaker | 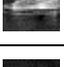 | sandal | 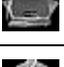 | bag | 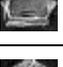 | boot |
| 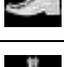 | boot | 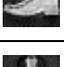 | sneaker | 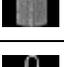 | shirt | 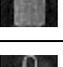 | T-shirt |
| 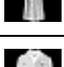 | dress | 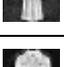 | bag | 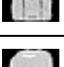 | bag | 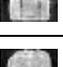 | sandal |
| 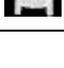 | coat | 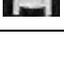 | pullover | 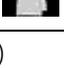 | T-shirt | 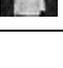 | shirt |

(b)

Fig 4. Examples of adversarial attack results on an ONN with cascaded MZIs for (a)number digit image classification; (b)fashion product image classification (left: image; right: classification result)

It can be observed that the attacked images are visually very similar to the original ones. They appear to be contaminated with noise very slightly. But it shall be noticed that in this task the "noise" is optimized, in comparison to random noise. The classification results for the original images and the attacked images are identical from human eyes. However, the actual classification results of the attacked images by using the ONN with cascaded MZIs are all incorrect, different from those of the original images. For two pairs of original images and attacked images, it can be observed that the maximum elements in the output vectors are changed, shown in Fig. 5. The ONN system is "cheated" and "fooled" by most of our attacked input images.

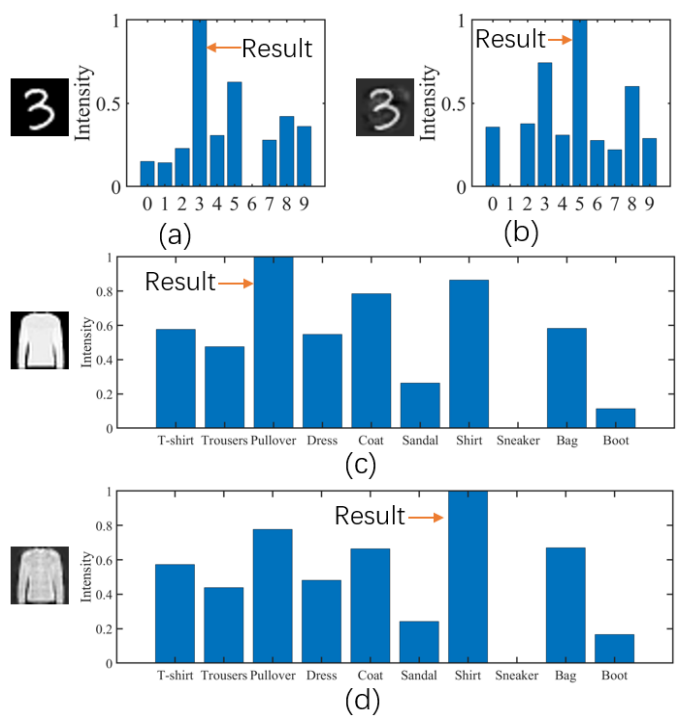

Fig. 5. Change of maximum element in the output vector (normalized) for a number digit image : (a)original image; (b)attacked image; and for a fashion product image : (a)original image; (b)attacked image;

The classification results after attack shown above are not pre-defined. As a more difficult task, a "selective" adversarial attack is further implemented. Each image is attacked for all of the nine other categories. The parameter $\gamma$ is set to be 1 and other parameters remain the same. The successful rate of attack is reduced to 61.87% (for MNIST) and 50% (for Fashion-MNIST). Some examples of selective attacking results for MNIST number digit images are shown in Fig. 6.

|  | Image | Result | Image | Result | Image | Result | Image | Result |
|---|---|---|---|---|---|---|---|---|
| Original | 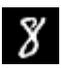 | 8 | 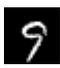 | 9 | 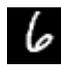 | 6 | 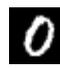 | 0 |
| Attack for 0 | 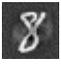 | 8 | 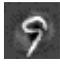 | 0 | 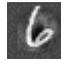 | 0 | NA | NA |
| Attack for 1 | 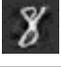 | 1 | 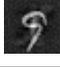 | 3 | 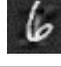 | 6 | 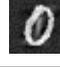 | 0 |
| Attack for 2 | 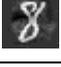 | 2 | 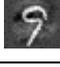 | 2 | 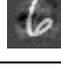 | 2 | 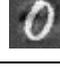 | 2 |
| Attack for 3 | 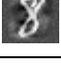 | 3 | 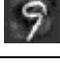 | 3 | 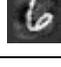 | 3 | 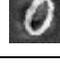 | 3 |
| Attack for 4 | 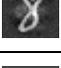 | 8 | 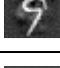 | 4 | 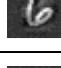 | 4 | 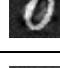 | 0 |
| Attack for 5 | 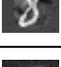 | 5 | 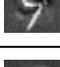 | 5 | 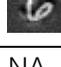 | 5 | 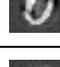 | 5 |
| Attack for 6 | 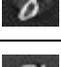 | 8 | 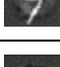 | 6 | NA | NA | 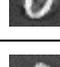 | 6 |
| Attack for 7 | 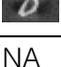 | 7 | 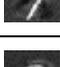 | 7 | 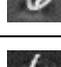 | 7 | 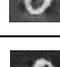 | 0 |
| Attack for 8 | NA | NA | 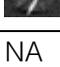 | 8 | 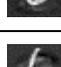 | 6 | 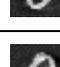 | 8 |
| Attack for 9 | 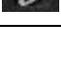 | 9 | NA | NA | 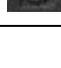 | 6 | 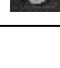 | 0 |

Fig. 6. Some examples of "selective" attacking results with different targets for MNIST number digit images (black: success; red: failure)

It can be observed that the original image can be transformed to another image with minor changes that will be mis-classified as the corresponding pre-defined one of the remaining nine categories, despite the failure cases. The differences between original images and attacked images in Fig. 6 are slightly higher than those in Fig. 4.

The adversarial attack can substantially degrade the classification performance of an ONN system by generating incorrect output results from apparently similar input images. This phenomenon is significant for both digital and optical machine learning systems, The investigation of adversarial attacks on linear classifiers based on vector-matrix multiplication also applies to other similar optical machine learning systems [21-24], in addition to ONN with cascade MZIs. In future works, the robustness of an ONN can be further enhanced against these adversarial attacks.